\documentclass[twocolumn,aps,prl,floatfix,superscriptaddress]{revtex4}
\usepackage{graphicx}
\usepackage{bm}
\usepackage{amsfonts}
\usepackage{amssymb}
\usepackage{amsmath}    
\def\buk{{\hat {\bm u}}_{\bm k}}
\def\bup{{\hat {\bm u}}_{\bm p}}
\def\buq{{\hat {\bm u}}_{\bm q}}

\def\ukp{u^+_{\bm k}}
\def\ukm{u^-_{\bm k}}

\def\bk{{\bm k}}
\def\bp{{\bm p}}

\def\bq{{\bm q}}
\def\bu{{\bm u}}
\def\bx{{\bm x}}
\def\bz{{\bm z}}
\def\bw{{\bm \omega}}
\def\hm{{\bm h}^-_{\bm k}}
\def\hp{{\bm h}^+_{\bm k}}
\def\hpm{{\bm h}^\pm_{\bm k}}

\usepackage{color}

\begin{document}
\title{Disentangling the triadic interactions in Navier-Stokes equations\footnote{Postprint version of the manuscript published in Eur. Phys. J. E {\bf 38}, 114 (2015)}}
\author{Ganapati Sahoo} 
\author{Luca Biferale} 
\affiliation{Department of Physics \& INFN, University of Rome Tor Vergata, Via della Ricerca Scientifica 1, 00133 Rome, Italy.}
\date{\today}
\begin{abstract} 
We study the role of helicity in the dynamics of energy transfer in a
modified version of the Navier-Stokes equations with explicit breaking
of the mirror symmetry.  We select different set of triads participating
in the dynamics on the basis of their helicity content. In particular,
we remove the negative helically polarized Fourier modes at all
wavenumbers except for those falling on  a localized shell of
wavenumber,  $|{\bk}| \sim k_m$.  Changing $k_m$ to be above or below
the forcing scale, $k_f$, we are able to assess the energy transfer of
triads belonging to different interaction classes.  We observe that when
the negative helical modes are present only at  wavenumber smaller than
the forced wavenumbers, an inverse energy cascade develops with an
accumulation of energy on  a stationary helical condensate. Vice versa,
when negative helical modes are present only at wavenumber larger  than
the forced wavenumbers, a transition from backward to forward energy
transfer is observed in the regime  when the minority modes become
energetic enough.  
\end{abstract} 

\maketitle 

\section{Introduction} 

It is known that the energy in a three dimensional homogeneous and isotropic
turbulent flow cascades forward, from the forcing scales to the dissipative
scales~\cite{k41}. When Reynolds number is high enough, an intermediate range
of scales develops where the energy  flux is constant~\cite{frisch}.  However,
systems like rotating flows~\cite{mininni,deusebio}, flows confined along one
direction~\cite{celani} and flows of conducting materials~\cite{brandenburg}
show an inverse energy transfer toward  larger and larger  scales. As a result,
it is still not clear what are the internal dynamical mechanisms that trigger
the direction of the energy flux in fully developed turbulence.  In this paper,
we present a series of numerical experiments done on a modified version of the
three-dimensional Navier-Stokes equations where a subset of Fourier modes have
been removed.  There are many different ways to achieve a mode reduction, from
the usual Galerkin truncation of all modes with $|{\bk}| >k_{\rm max}$ to more
refined self-similar truncation done on a fractal-Fourier set~\cite{fractal}.
Here, we are interested to further explore the possibility to reduce mode on
the basis of their helicity content~\cite{waleffe,constantin,biferale2013}. Helicity,
together with energy,  is an inviscid invariant of three-dimensional
Navier-Stokes equation and it is known to play a key role both for
hydrodynamical and magnetohydrodynamical
systems~\cite{biferale-jstat,moffatt69,moffatt92,brissaud,laing,ditlevsen,holm,biferaleh,chen,chen2,dubrulle,biskamp,baer,mininni2010}.
In previous works~\cite{biferale2013,biferale2012} we have shown that by
constraining the velocity field  to develop fluctuations with only one sign of
helicity, the energy flows backward: from the forced scale to the largest scale
in the system, without reaching a steady state if not confined on a finite box
or without the addition of external friction. More recently, we
have shown that the inverse cascade regime, observed for  the fully helical
case,  is highly  fragile~\cite{sahoo2015}: it is enough to have a tiny number of helical modes
with the opposite sign distributed uniformly on the Fourier space to revert the
system to a forward cascade regime. Such a conclusion is also
supported by arguments based on absolute
equilibrium~\cite{herbert,kraichnan}. In this paper we explore the case
when all Fourier modes have the same helicity (say positive) except for a small subset
possessing also the opposite (negative) helicity. The latter being limited to
belong to a tiny  shell in Fourier space. The goal is to make a further step
toward a better understanding of the dynamics of energy transfer in
Navier-Stokes equations, triad-by-triad. The paper is organized as follows:  In
Sec. 2 we briefly describe helical decimation and write the Navier-Stokes
equations for helical Fourier modes; In Sec. 3 we discuss the results from our
direct numerical simulations for two different series of computations, either
when the negative helical modes are confined to a shell of wavenumbers larger
than the force scale or in the opposite case. Conclusions can be found in  Sec.
4.

\section{Helically decomposed Navier-Stokes equations}

\begin{figure*}[!htb]
\center
  \includegraphics[scale=0.43]{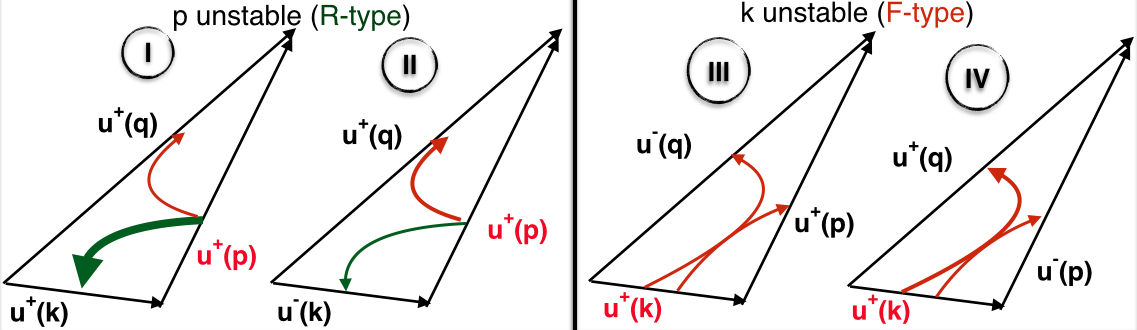}
  \caption{(Color online) Schematic presentation of triads~\cite{waleffe}:
Triads where the two largest wavenumbers have the same sign of helicity are
responsible for a reverse transfer of energy  and are called of R-type.  They
include triads of Class I and of Class II.  Triads where the two  largest
wavenumbers have opposite sign of helicity are responsible for forward transfer
of energy and are called of F-type. They include triads in Class III  and
Class IV. For R-type (F-type) the Fourier mode with the  medium (smallest)
wavenumber is  unstable  and transfers energy to the other two
Fourier modes. The arrows (green for reverse and red for forward) show
direction of energy transfer. }
  \label{fig1}
\end{figure*}

The velocity field in a periodic domain is expressed by the Fourier series
\begin{align}
\bu(\bx) = \sum_{\bk} \buk e^{i\bk\cdot\bx},
\end{align}
 where the modes $\buk$ satisfy the incompressibility condition
$\bk\cdot\buk=0$ and can be exactly decomposed in terms of the helically
polarized waves as~\cite{waleffe,constantin}
\begin{equation}
\label{eq:dec}
  \buk  = \ukp \hp +\ukm \hm.
\end{equation}
The eigenvectors of the curl $\hpm$ are given by
\begin{align}
\hpm = \hat{\nu}_{\bm k} \times
\hat{k} \pm i \hat{\nu}_{\bm k},
\end{align}
so that $i {\bk} \times \hpm = \pm k \hpm$; where $ \hat{\nu}_{\bm k}$ is an
unit vector orthogonal to ${\bk}$ such that $\hat{\nu}_{\bm k} = -
\hat{\nu}_{-\bk}$ to enforce reality of the field. One can choose for example~\cite{waleffe}:  
\begin{align}
\hat{\nu}_{\bk} = \frac{{\bz} \times {\bk}}{ || {\bz} \times {\bk} ||},
\end{align}
 where $\bz$ is any arbitrary vector.  The eigenvectors $\hpm$
satisfy the orthogonality condition ${\bm h}^{s}_\bk\cdot{\bm
h}^{t*}_\bk=2\delta_{st}$, where $s,t = \pm$  denote the signs of the helicity 
and $*$ is for  the complex conjugate.
We define a projector 
\begin{align}
  \label{eq:poperator}
  {\mathcal P}^\pm_{\bk} \equiv \frac {\hpm \otimes {\bm h}^{\pm*}_\bk} {{\bm h}^{\pm*}_\bk \cdot \hpm},
\end{align}
which projects the Fourier modes of the velocity on eigenvectors $\hpm$
as
\begin{align}
  \label{eq:projection}
  {\mathcal P}^\pm_{\bk} \buk = {\hat \bu}^\pm_\bk = u^\pm_\bk\hpm.
\end{align}
The Navier-Stokes equations can be decomposed in terms of the evolution of  
velocities with positive or negative sign of helicity as follows:
\begin{equation}
\label{eq:NS}
\frac{\partial\bu^\pm(\bx)}{\partial t} + {\cal D}^\pm {\rm \bf N}[\bu(\bx),\bu(\bx)] = \nu\nabla^2\bu^\pm(\bx)+{\bf f}^{\pm},
\end{equation}
where the operator  ${\cal D}^{\pm}(\bu)$ acts on a generic three-dimensional 
vector field by projecting all Fourier components on  $\hpm$:
\begin{equation} 
  \label{eq:projector}
  {\cal D}^{\pm}{\bu}(\bx) \equiv \sum_{\bk} e^{i\bk\cdot\bx}\,{\mathcal P}^{\pm}_{\bk} {\buk},  
\end{equation}
and   ${\rm \bf N}[\bu(\bx),\bu(\bx)]$ is the nonlinear terms of the Navier-Stokes equations~\cite{biferale2012}. 
The total energy and the total helicity can also be easily expressed in terms of the helical modes: 
\begin{align}
    E &= \int d^3 x \, |\bu(\bx)|^2 = \sum_{\bk} |\ukp|^2 + |\ukm|^2,\\
    H &= \int d^3 x \, \bu(\bx) \cdot \bw(\bx) = \sum_{\bk} k(|\ukp|^2 - |\ukm|^2),
\end{align} 
where $\bw(\bx)=\nabla\times\bu(\bx)$ is the vorticity. From the above
expression one can introduce the energy spectrum for positive and for negative
helical modes~\cite{chen,chen2}: 
\begin{align}
E^+(k) =  \sum_{|\bk| \in [k,k+1]} |\ukp|^2; \\
E^-(k) =  \sum_{|\bk| \in [k,k+1]} |\ukm|^2.
\end{align}
Plugging the decomposition (\ref{eq:dec}) in to the Navier-Stokes equations (\ref{eq:NS}) it is easy to
realize that the nonlinear term consists of triadic interactions with eight
(four for the evolution of $u^+$ and four for the evolution of $u^-$) possible
helical combinations of  the generic modes $u^{s_\bk}_\bk$, $u^{s_\bp}_\bp$,
$u^{s_\bq}_\bq$ forming an interacting triad, i.e., $\bk+\bp+\bq=0$, for
$s_\bk=\pm$, $s_\bp=\pm$, $s_\bq=\pm$~\cite{waleffe} (see fig.~\ref{fig1} where
for  simplicity we assume that $k\le p \le q$). The four classes of triads are
classified as follows: Class I, containing triads formed with all wavenumbers
having the same sign of helicity, i.e., $(+, +, +)$; Class II, made of  triads
where the two smallest wavenumbers have opposite sign of helicity and the two
largest wavenumbers have the same sign of helicity, i.e., $(-, +, +)$; Class III,
containing triads where the two smallest wavenumbers have the  same sign of
helicity and the two largest wavenumbers have an opposite sign of helicity,
i.e., $(+, +, -)$; and Class IV, made of  triads where the two smallest
wavenumbers and the two largest wavenumbers have opposite sign of helicity,
i.e., $(+, -, +)$ (see fig.~\ref{fig1}).  In ~\cite{waleffe}, studying the
instability of the energy exchange among modes of each single triad,  it was
argued that the triads in Class III and Class IV transfer energy from the smallest
wavenumber to the other two wavenumbers and are responsible for the forward
cascade of energy.  Whereas for the triads in Class I and Class II, the Fourier mode with the
medium wavenumber transfers energy to the other two Fourier modes.  These sets
of triads might then contribute to both forward and backward energy transfers.
The presence of competing interactions might suggest that the direction of the
energy transfer mechanism is not set {\it a priori}.  Depending on the empirical
realization (the forcing scheme, the boundary conditions, the coupling with
other active dynamical fields as for the case of conducting 
flows~\cite{celani,brandenburg,alexakis14,alexakis15}) different directions of the energy could be
developed.  As said, in the whole system where all triads are present and with
a neutral homogeneous and isotropic external forcing,  energy is observed to be
transferred forward: from large to small scales. On the other hand, a system in
which only modes of one sign of helicity are present, i.e., the dynamics is
restricted to interacting triads with $s_\bk=s_\bp=s_\bq$ (Class I), energy
cascades from small scales to the large scales~\cite{biferale2012}.  This was
reconducted to the fact that helicity becomes a sign-definite quantity for such
subset of interactions.  In a recent work it was observed~\cite{sahoo2015} that
presence of few percent of modes with opposite sign of helicity changes the
direction of energy transfer in a singular manner: a few modes with both sign
of helicity at all scales, even though one type is a small fraction of other
type, allows formation of triads with two largest wavenumbers having opposite
signs of helicity which efficiently transfers energy to the small scales.

\begin{figure}[!htb]
  \includegraphics[scale=0.6]{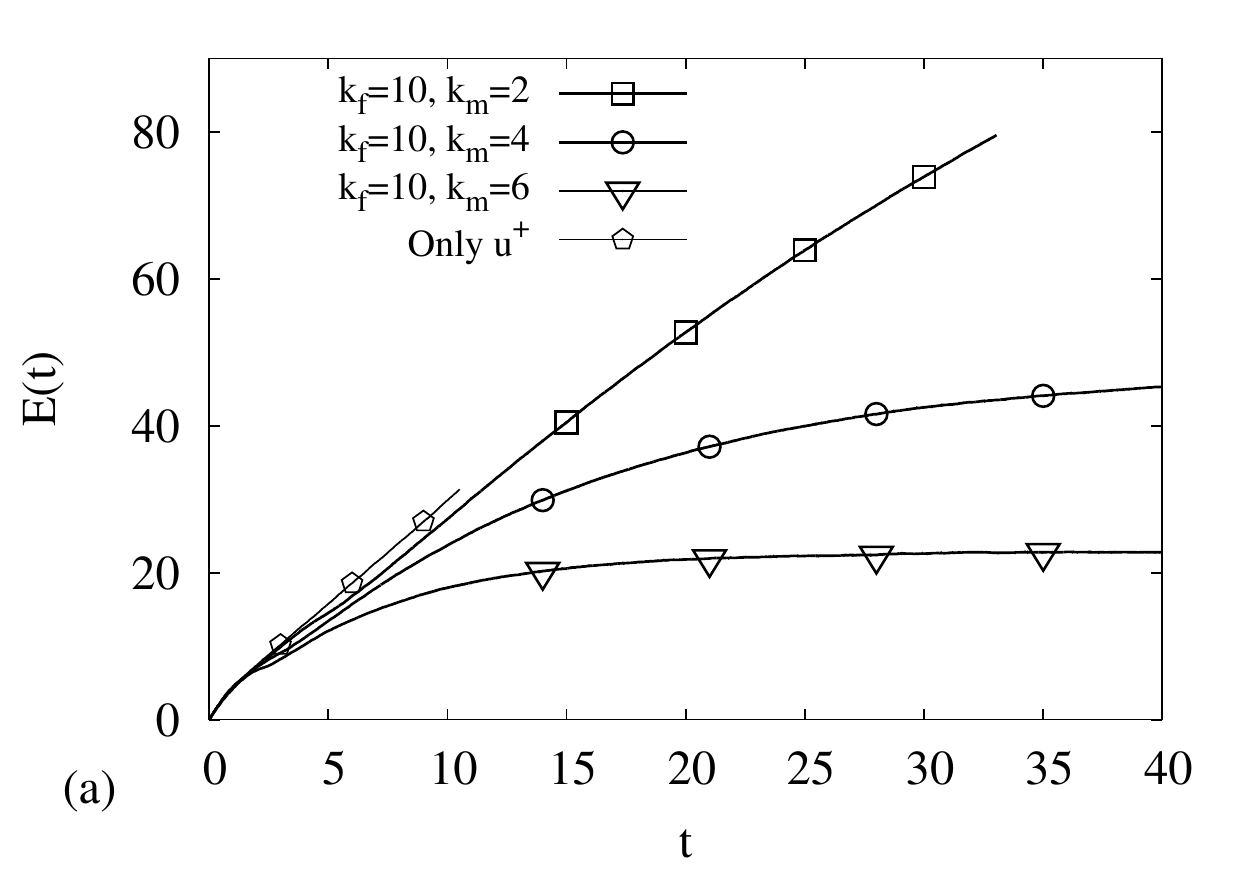}\\
  \includegraphics[scale=0.6]{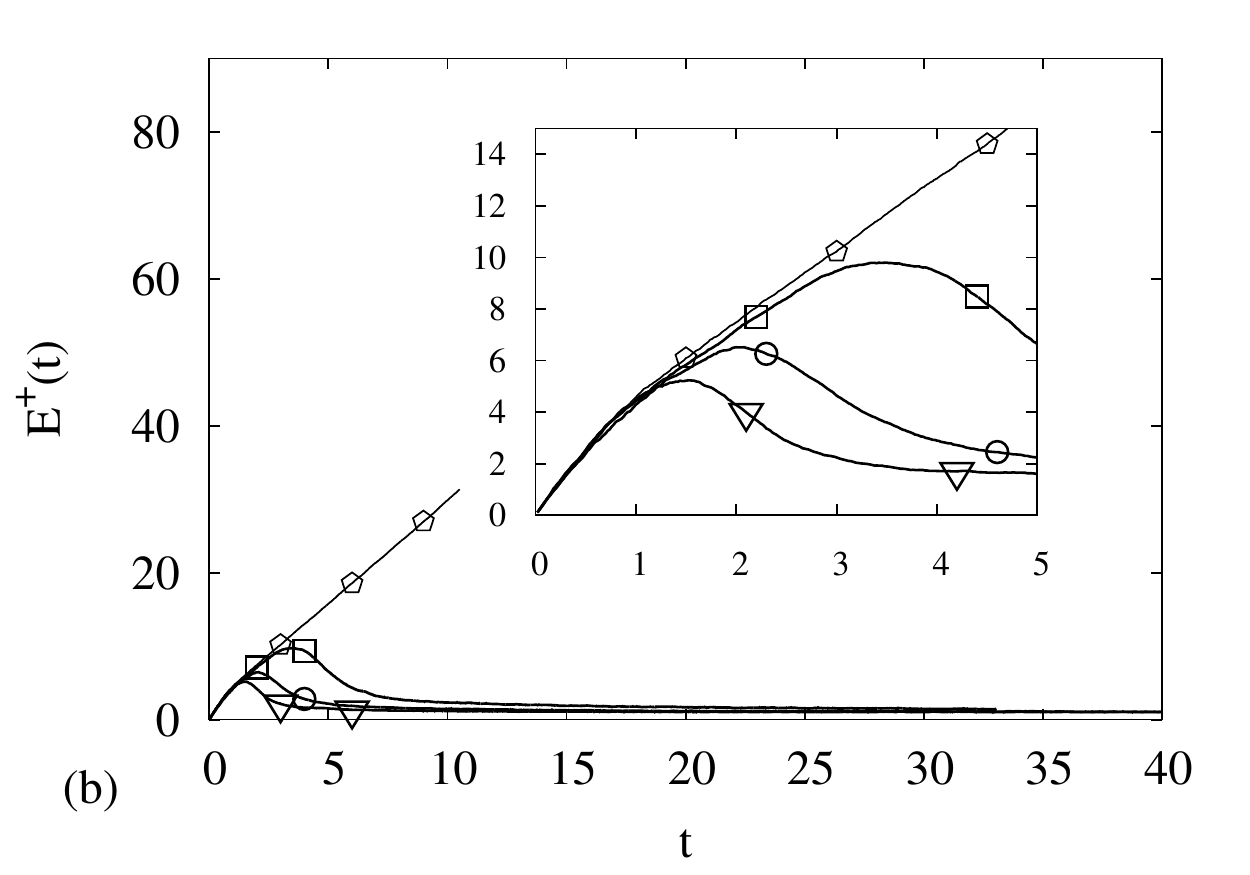}\\
  \includegraphics[scale=0.6]{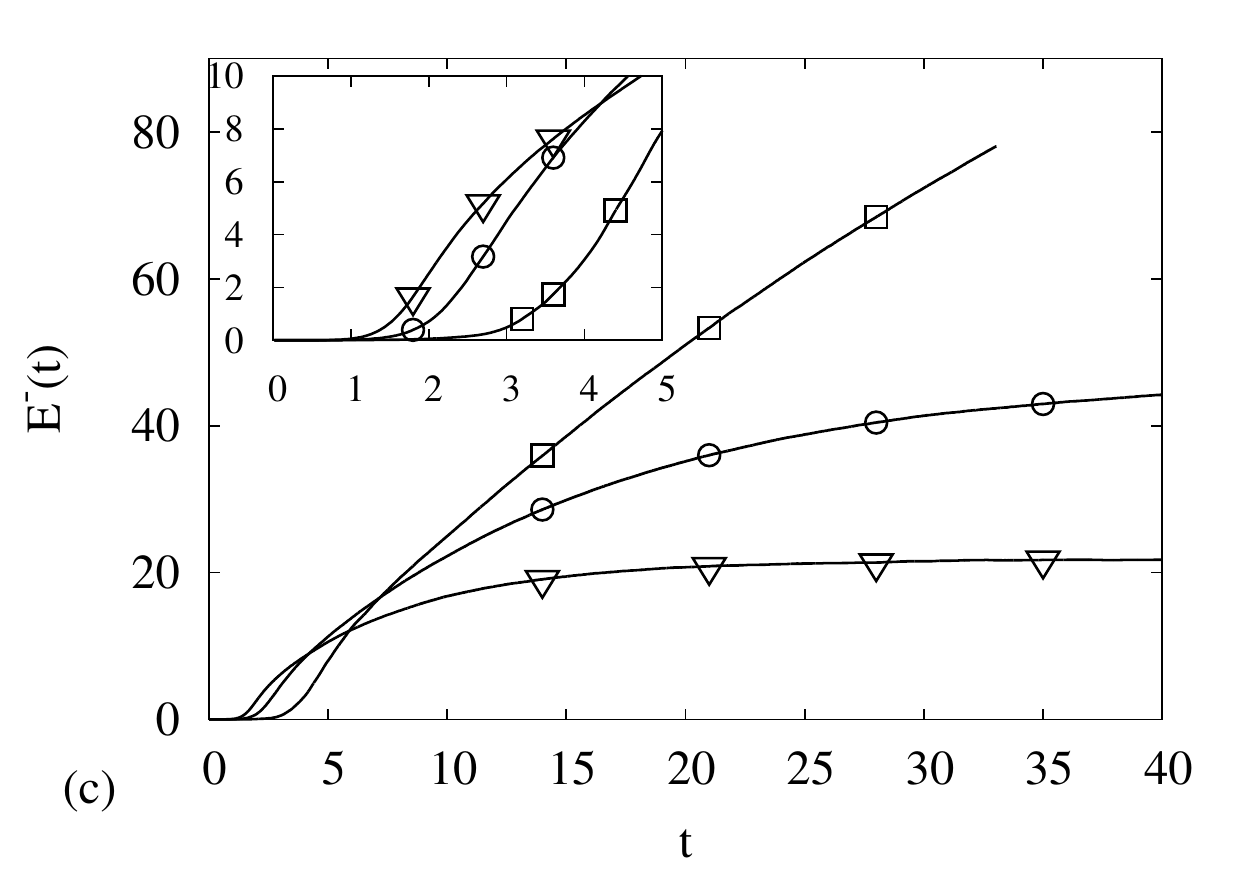}
  \caption{Time evolution of the  energy based on  (a)  all modes, (b) only
positive helical modes, and (c) only negative helical modes, for the three
cases where $k_f\in [10,12]$ and $\ukm = 0$ except wavenumbers around
$k_m=2,4,6$.  In the insets of panel (b) and (c) we show an enlargement of the
initial period.  Notice that the dynamics is first dominated by the sucking of
energy by  the positive helical modes at low wavenumbers and then it switches
to transfer energy only to the negative ones. In panels (a) and (b) we also
show the results for the growth of energy when only positive helical modes are
present. In the latter case the growth of the energy in the positive helical
modes is not stopped.  }
  \label{fig2}
\end{figure}

 In this work we attempt to control the energy transfer mechanism in presence
of two different set of triads (Class II and Class IV).  To do that we remove
the negative helical modes for all wavenumbers, except for those falling in one
shell of Fourier modes $|{\bk}| \in D_m$ with $ D_m \equiv \{ {\bk} : |{\bk}| 
\in [k_m,k_{m}+1] \}$. We consider two cases (i) $k_m <k_f$ and (ii) $ k_m >k_f$,
where $k_f$ is the typical wavenumber where we apply the external forcing
mechanism.  To do that  we define an operator ${\cal D}_m$ which projects the
velocity on $\hp$ for wavenumbers outside the coset of
$D_m$:
\begin{equation} 
  \label{eq:projv}
  \bu'(\bx) \equiv {\cal D}_m {\bu}(\bx) \equiv \sum_{\bk} e^{i{\bm k}\bx}\, [(1-\gamma_{\bk}) + \gamma_{\bk} {\mathcal P}^+_{\bk}] {\buk},  
\end{equation}
where $\gamma_{\bk}=0$ for $\bk \in D_m$  and $\gamma_{\bk}=1$ otherwise. 
 The decimated three-dimensional Navier-Stokes equations are given by:
\begin{equation} 
  \label{eq:ns+++}
  \partial_t \bu' = {\cal D}_m [- (\bu' \cdot \nabla) \bu' -{\bm \nabla} p'] +\nu \Delta \bu' + {\bf f}', 
\end{equation}
where $p'$ is the  pressure, $\nu $ is the viscosity and ${\bf f}'$ is the external forcing (see later for details). 
Although nonlinear terms of the decimated Navier-Stokes equations
do not have Lagrangian properties~\cite{moffatt14}, it can still be shown that 
both  energy 
\begin{align}
 E &= \sum_{\bk} ( |\ukp|^2 + (1-\gamma_{\bk})|\ukm|^2), \label{eq:ealpha}
\end{align}
 and helicity
\begin{align}
 H &= \sum_{\bk} k( |\ukp|^2 - (1-\gamma_{\bk})|\ukm|^2), \label{eq:halpha}
\end{align}
are invariants of eq.(\ref{eq:ns+++}) in the inviscid and unforced limit. 
\begin{table}
  \begin{center}
  \begin{tabular}{| c | c | c | c | c | c | c | c |}
  \hline
   RUN & $N$   & $L$    & $k_f$     & $k_m$ & $\nu$   & $\delta t$ & $F_0$ \\
  \hline
   R1  & $512$ & $2\pi$ & $[10,12]$ & $6$   & $0.002$ & $0.0001$   & $5$ \\ 
   R2  & $512$ & $2\pi$ & $[10,12]$ & $4$   & $0.002$ & $0.0001$   & $5$ \\ 
   R3  & $512$ & $2\pi$ & $[10,12]$ & $2$   & $0.002$ & $0.0001$   & $5$ \\ 
   R4  & $512$ & $2\pi$ & $[4,6]$   & $10$  & $0.001$ & $0.0001$   & $5$ \\ 
   R5  & $512$ & $2\pi$ & $[4,6]$   & $16$  & $0.001$ & $0.0001$   & $5$ \\
  \hline
  \end{tabular}
  \end{center}
  \caption{$N$: number of collocation points along each axis. $L$: size of
the simulation box. $k_f$: range of forced wavenumbers. $k_m$: 
wavenumber of the shell with also negative helical modes. $\nu$: kinematic viscosity.
$\delta t$: time step. $F_0$: forcing amplitude.}
  \label{table1}
\end{table}
\begin{figure*}[!htb]
  \includegraphics[scale=0.7]{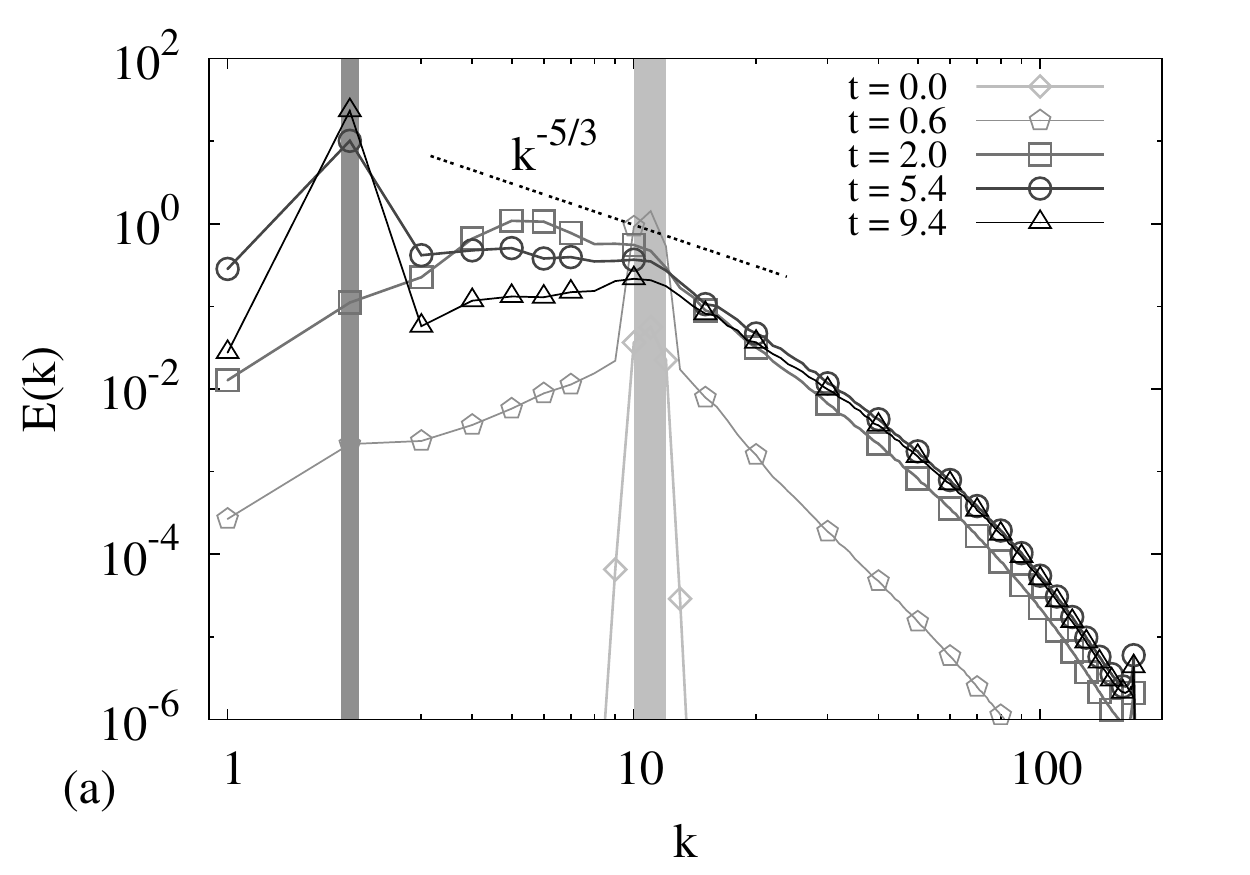}
  \includegraphics[scale=0.7]{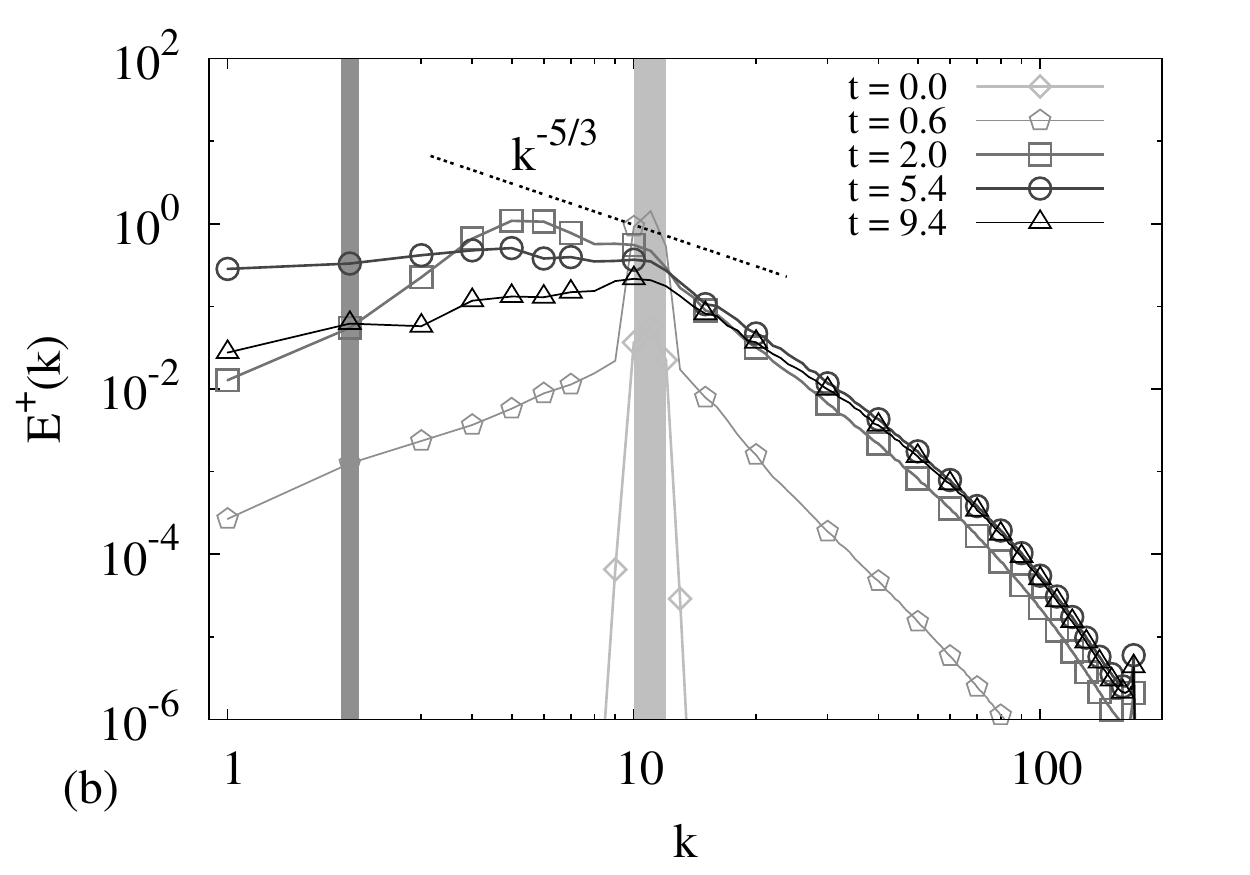}\\
  \includegraphics[scale=0.7]{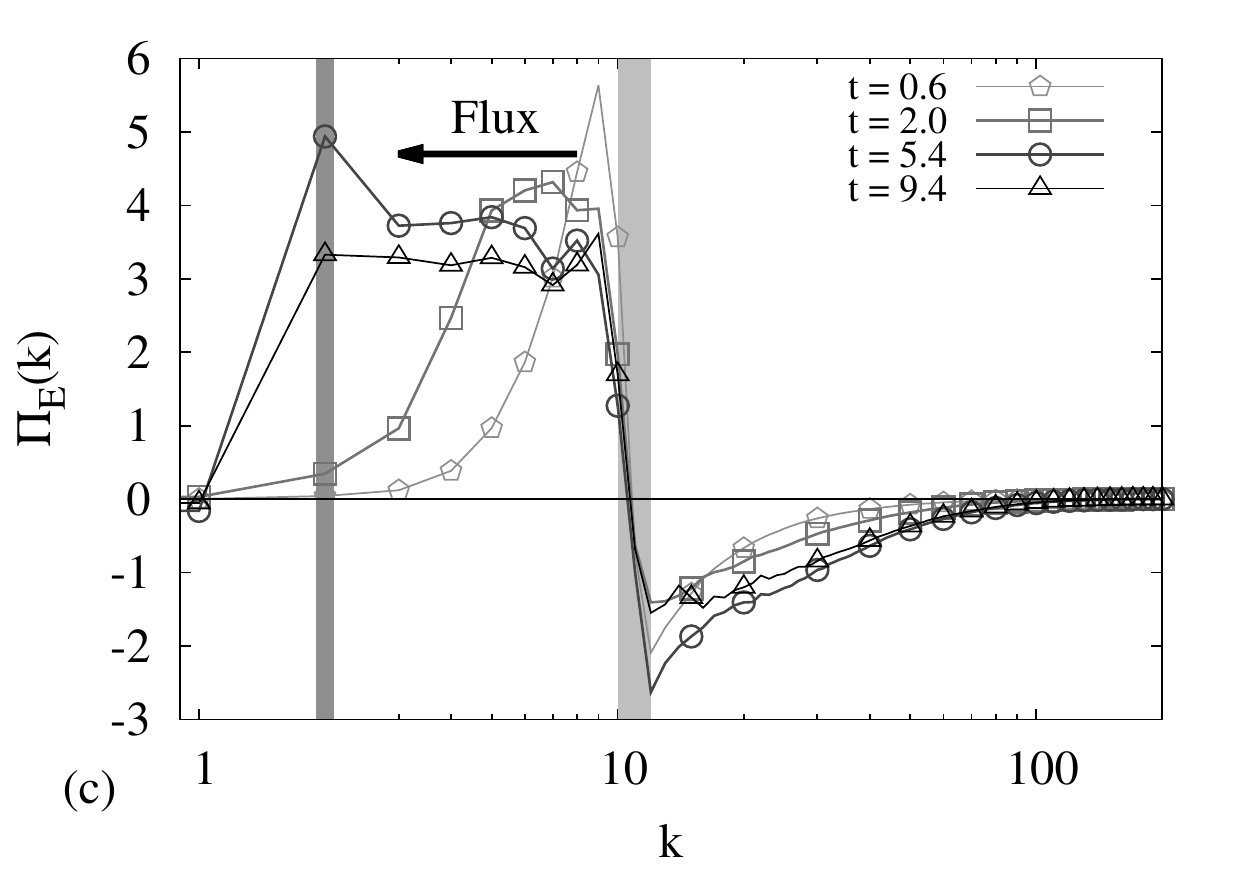}
  \includegraphics[scale=0.7]{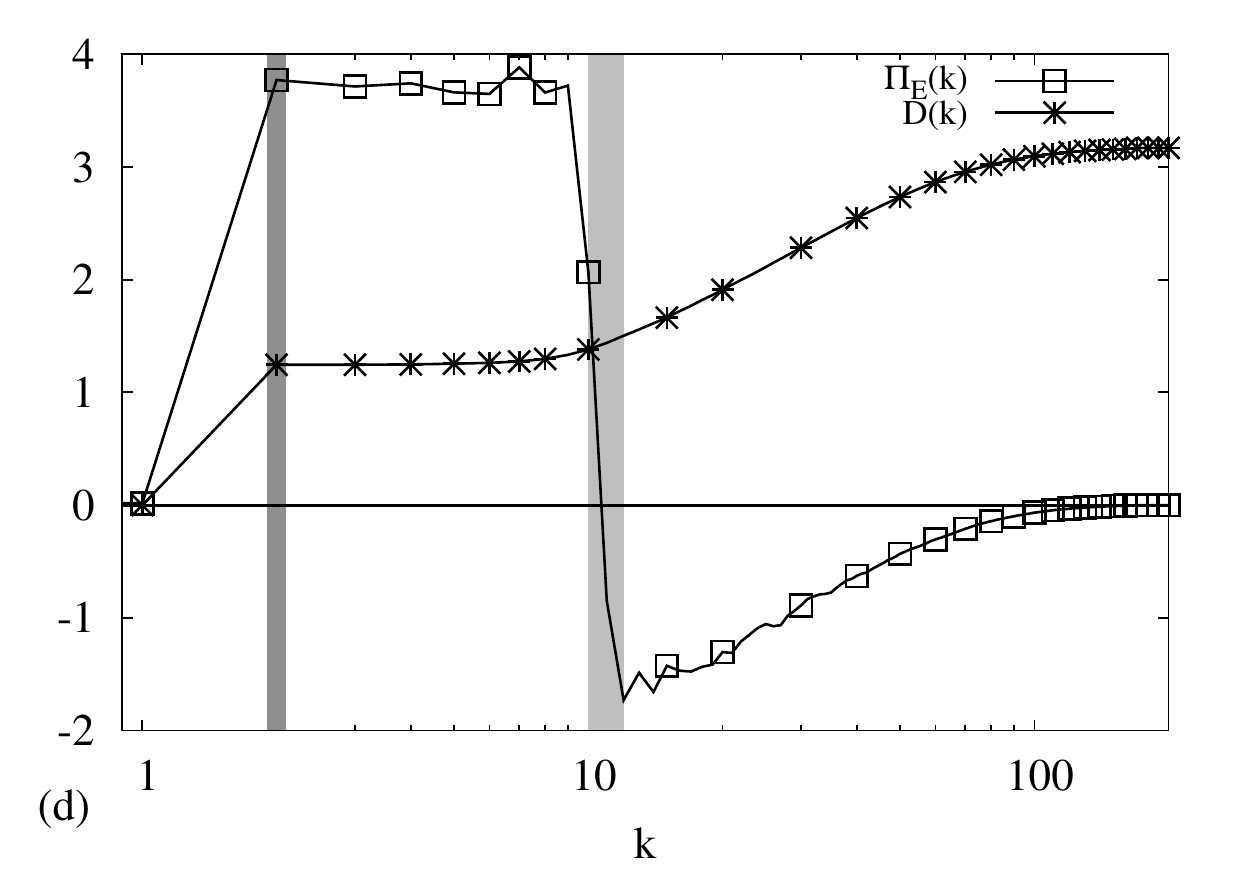}
  \caption{(a) Log-log plot of total energy spectra at different times. (b) The
same of (a) for the positive helical modes spectrum. The mismatch between the
two spectra for $k=k_m$ is due to the energy of the negative helical modes. We
have drawn a dashed line with slope of -5/3 to highlight the possible growth of
inverse cascade spectrum when there is a large inertial range of scales. (c)
Fluxes of energy (see definition (\ref{eq:flux})). 
(d) Comparision of energy flux ${\rm \Pi_E}(k)$ and dissipation
${\rm D}(k)$ (see text) at the time when the simulation is stopped ($t\sim32$,
see fig.~\ref{fig2}).  The forced wavenumbers at $k_f\in [10,12]$ are marked
with a light grey band, while  the wavenumbers with negative helical modes
around $k_m=2$ are in dark grey.}
  \label{fig3}
\end{figure*}
\begin{figure*}[!htb]
  \includegraphics[scale=0.7]{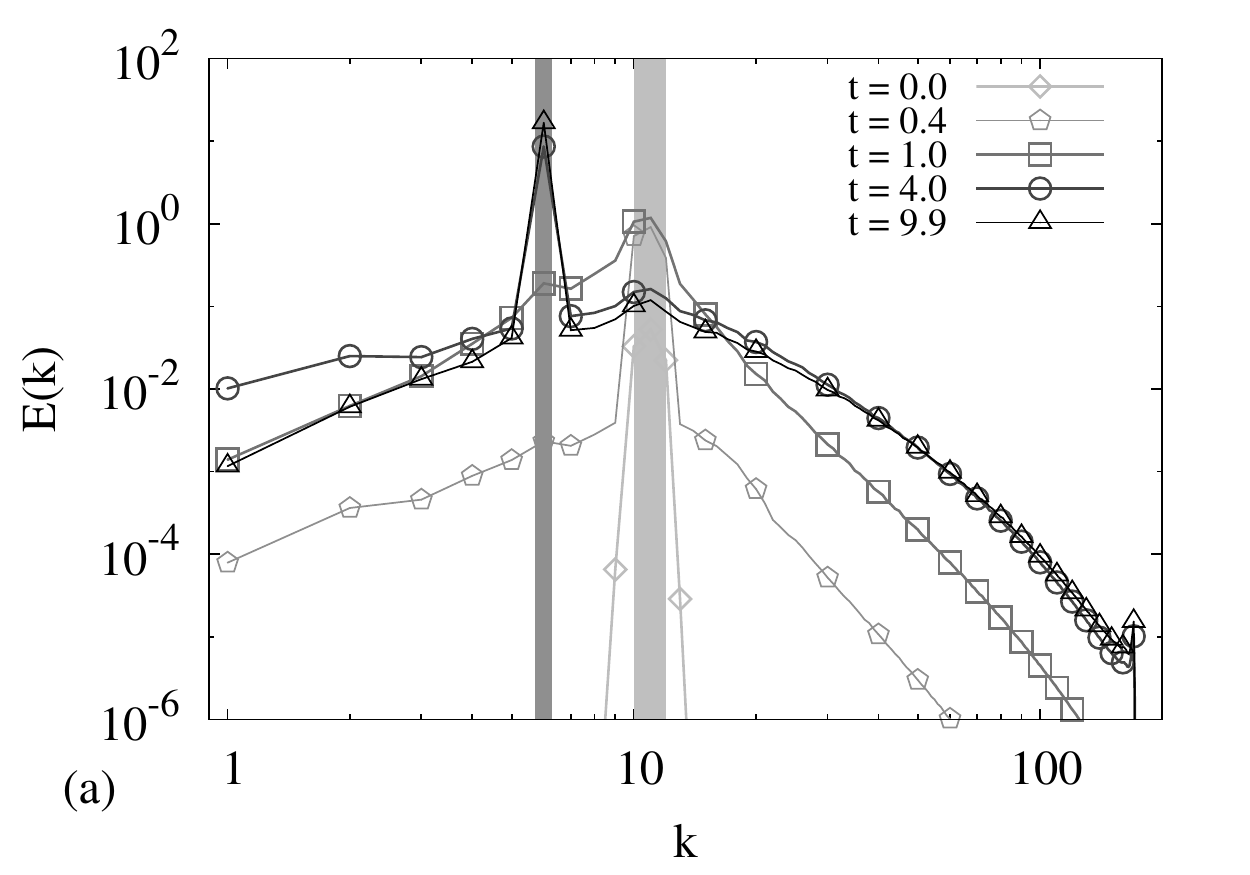}
  \includegraphics[scale=0.7]{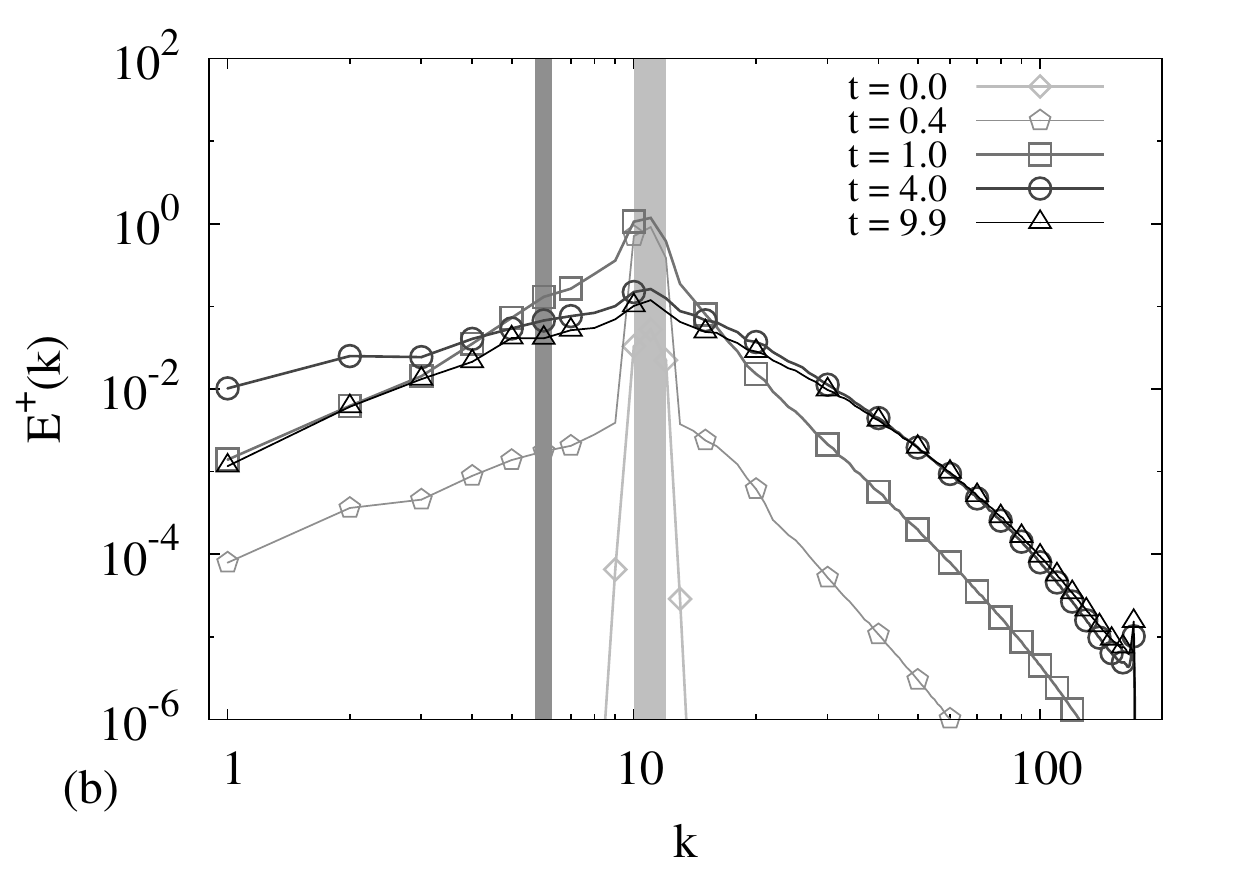}\\
  \includegraphics[scale=0.7]{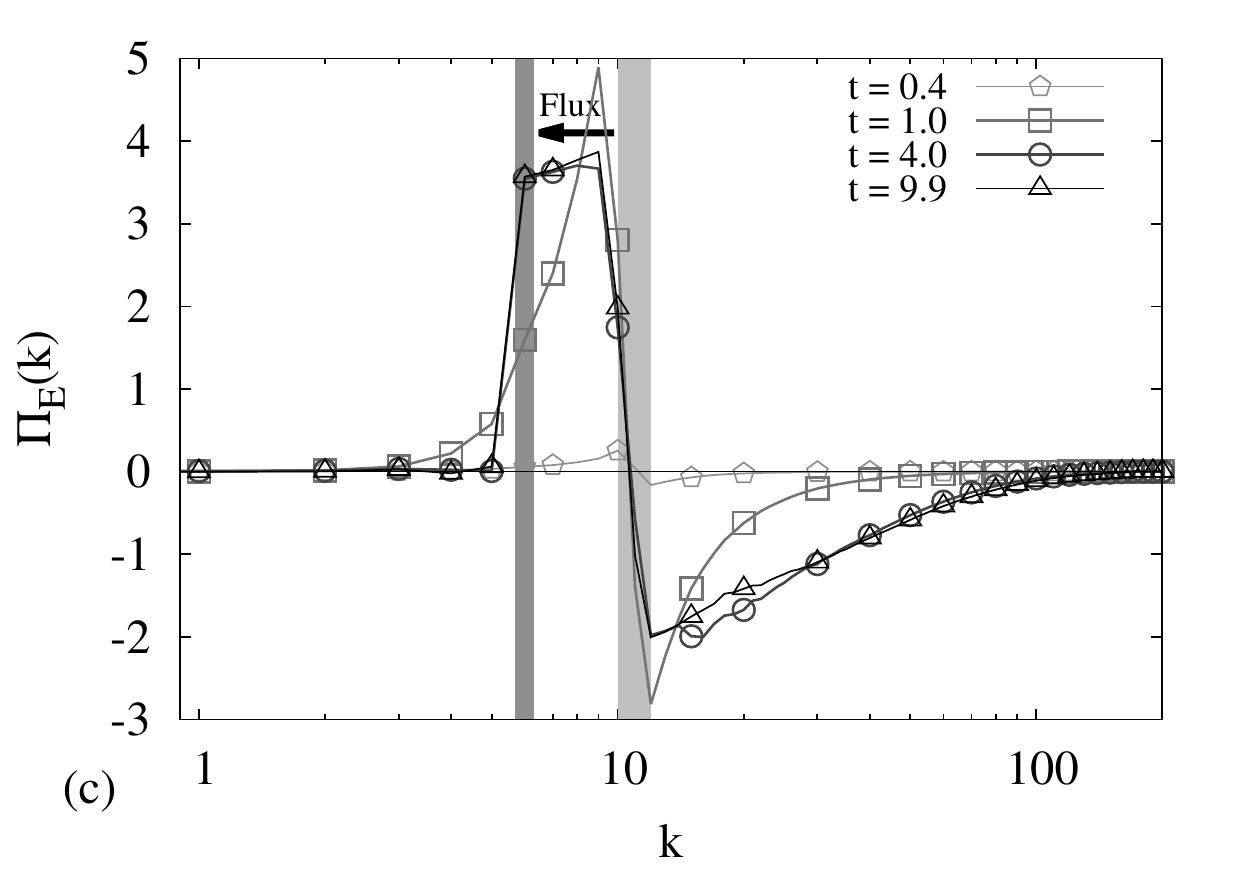}
  \includegraphics[scale=0.7]{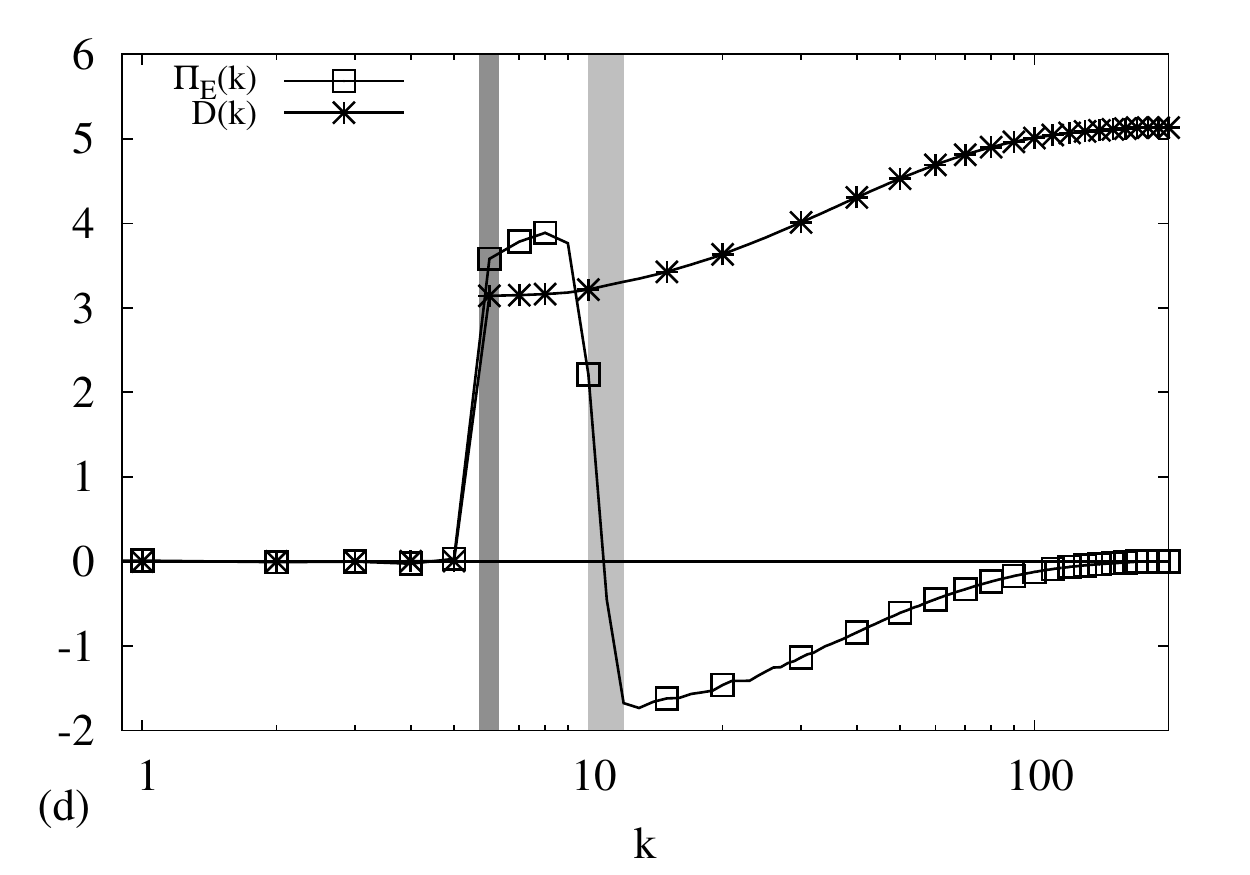}
  \caption{The same of fig.~\ref{fig3} but for the case when $k_m=6$,
except for (d) where energy flux and dissipation are compared at $t\sim40$ when
the simulation is stopped (see fig.~\ref{fig2}).}
  \label{fig4}
\end{figure*}

\section{Direct Numerical Simulations}
A pseudo-spectral spatial method is adopted to solve
eqs.~(\ref{eq:ns+++}), fully dealiased with the two-thirds rule;
time stepping is implemented with a second-order Adams-Bashforth
scheme. We performed different run up to a  resolution of $512^3$ collocation points, by changing the forced wavenumbers and the shell of modes where negative helical waves are retained.
We applied a random Gaussian force with
\[\langle f_i(\bk,t) f_j(\bq,t')\rangle = F(k) \delta(\bk-\bq) \delta(t-t')
Q_{ij}(\bk),\] where the projector $Q_{ij}(\bk)$ ensures incompressibility and
 $F(k) = F_0 k^{-3}$; the forcing amplitude $F_0$ is nonzero
 only for $ k_f \in [k_{\rm min}:k_{\rm max}]$.
Table.~\ref{table1} lists the details of various simulations.  Moreover,  we always projected the forcing on its positive helical components  in order to ensure maximal helicity
injection.  
We carried out two sets of simulations; First we retained the
negative helical modes in a shell of  wavenumbers $\sim k_m$ smaller than the forced
wavenumbers $k_f$, while  in the second case
we retained the negative helical modes at
a $k_m>k_f$. In the first set, negative helical modes exist only at
wavenumbers smaller than the forcing mechanisms so effectively we add triads of Class II to the triads of Class I.
In the second set, negative helical modes exist at higher wavenumbers,
resulting in the addition mainly of triads of Class III and Class IV.
\begin{figure}[!htb]
  \includegraphics[scale=0.7]{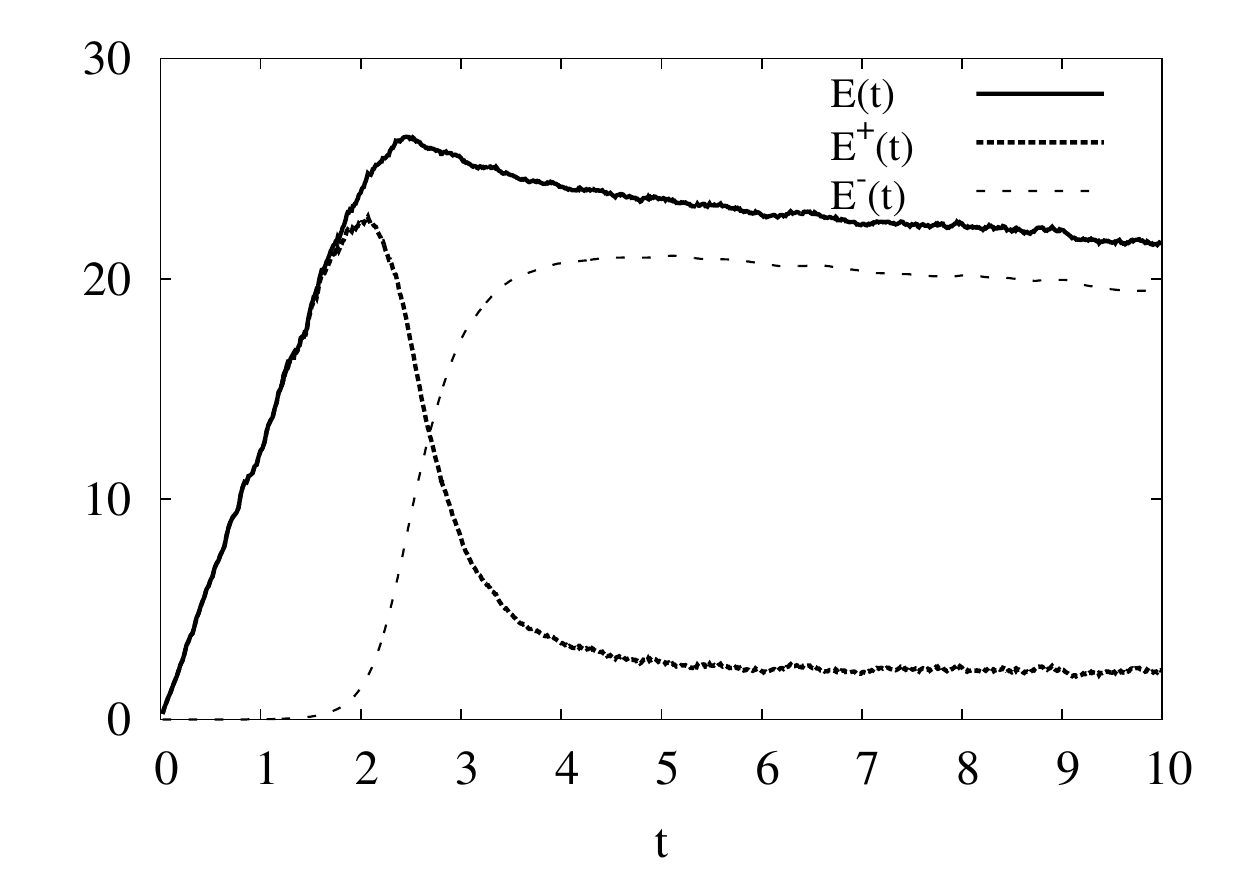}
  \caption{Time evolution of total energy $E(t)$, energy of positive
helical modes $E^+(t)$, and energy of negative helical modes $E^-(t)$ when $k_f\in [4,6]$ and $k_m = 16$.}
  \label{fig5}
\end{figure}
\begin{figure*}[!htb]
  \includegraphics[scale=0.7]{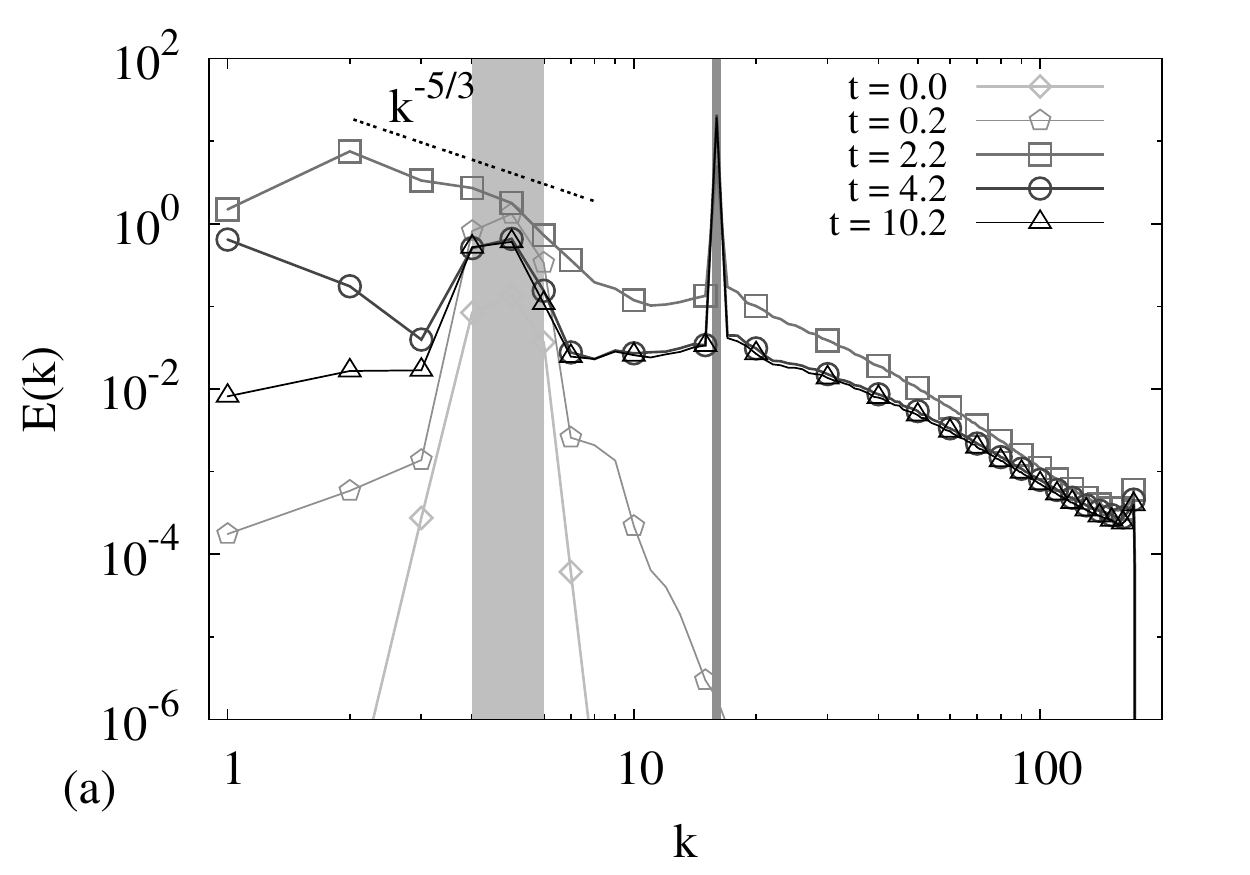}
  \includegraphics[scale=0.7]{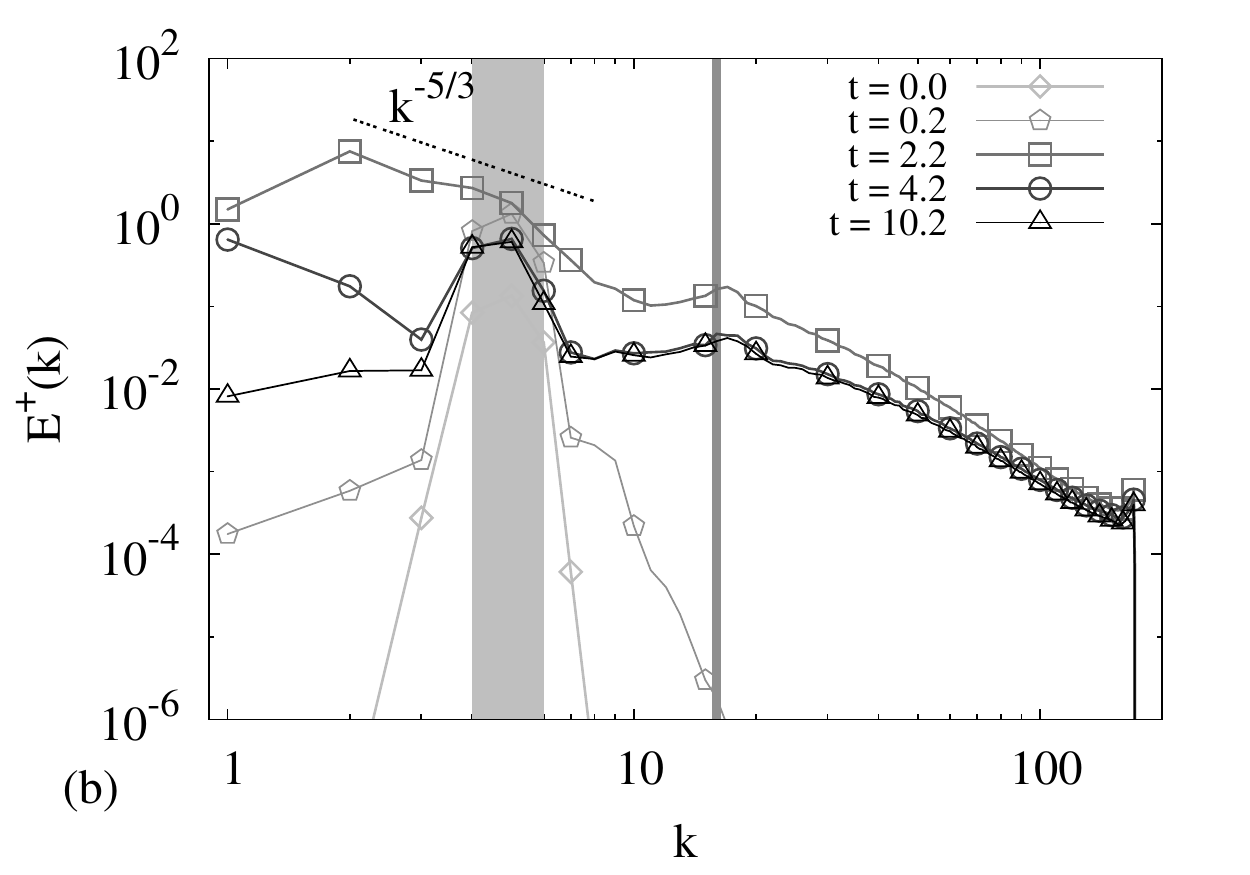}\\
  \includegraphics[scale=0.7]{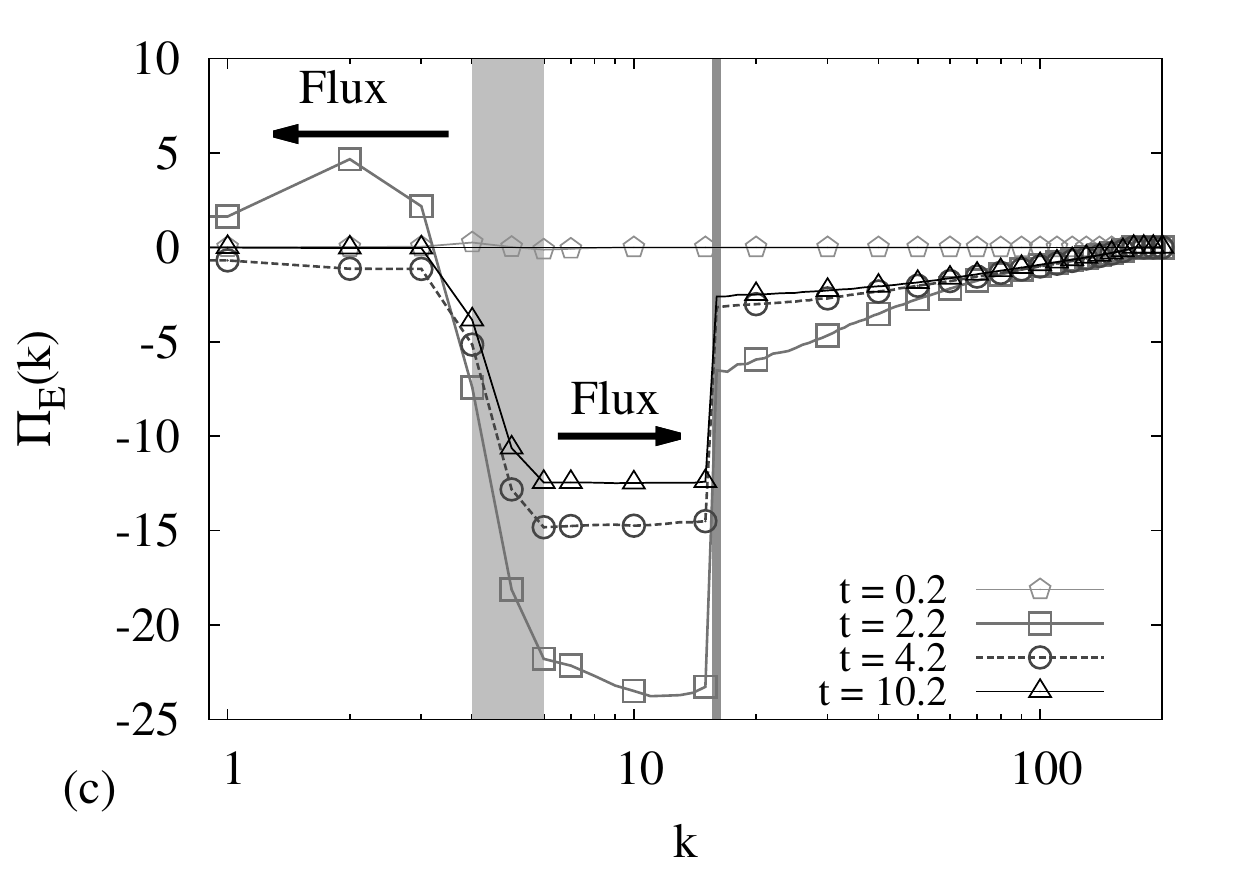}
  \includegraphics[scale=0.7]{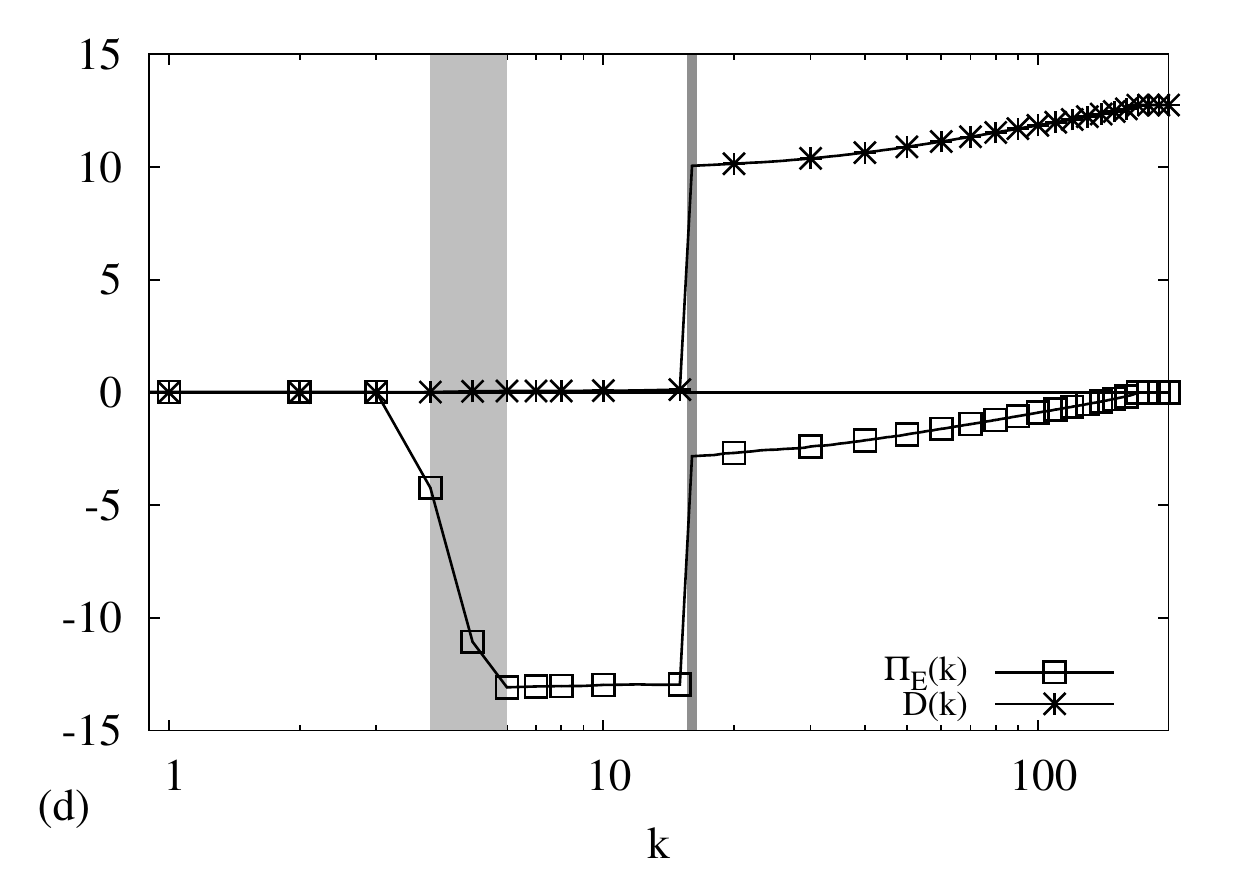}
  \caption{The same of fig.~\ref{fig3} but for the case when $k_f\in [4,6]$ and $k_m=16$,
except for (d) where energy flux and dissipation are compared at $t~10$ when
the simulation is stopped (see fig.~\ref{fig5}).}
  \label{fig6}
\end{figure*}

\subsection{Energy transfer for $k_m<k_f$}
In this set of simulations we keep $k_f\in[10,12]$ and change the value of
$k_m$ to $2$, $4$ and $6$.  Figure~\ref{fig2}(a) shows the evolution of energy
in the three cases. We always observe a steady inverse energy cascade which
reaches a statistically steady state, except for $k_m=2$ where the run was not
long enough to stabilize the system. Notice that we never introduced
an external energy sink at large scales. Therefore, a statistically stable
system means that  a stable large scale helical condensate is formed with an
energy large enough to be dissipated directly by molecular viscosity.  The
growth of energy in the positive and negative helical modes are shown
separately in fig.~\ref{fig2} (b) and (c). It is striking to note that in the
steady state the negative helical modes, existing only at $k=k_m$, carry almost all the
energy of the system, signaling that the inverse energy cascade process is
very efficient to move energy to the opposite helical modes via Class II
interactions. Moreover, the negative helical modes act as sinks and do not allow the
inverse cascade to proceed further to larger scales, stabilizing a condensate
to a given wavenumber, independent of the size of the box. A statistically
stationary state is then reached only when molecular drag becomes efficient at
such scales.  Initially the growth of energy is in the positive helical modes,
shown in the insets of panels (b) and (c). There is a critical change in the
dynamics of the system when the negative helical modes become energetic enough
(i.e., for the $k_m=2$ case around $t \sim 3$). The positive helical modes
at $k<k_m$ lose their energy as they form triads of Class III or Class IV with 
the negative helical modes and therefore contribute to the formation of condensate
 at $k\sim k_m$. To better understand the
dynamics among different wavenumbers we show the spectrum of energy at
different times in fig.~\ref{fig3}, for the case $k_m=2$. From
fig.~\ref{fig3}(a) we  see that at initial times ($t<2.0$) the growth of energy
in the large scales ($k<k_f$) is due to an inverse transfer to the positive
helical modes. This transfer is driven  by triads of Class I. When the negative
helical modes at $k=k_m$ becomes energetic enough ( $t \sim 5$)  the positive
helical modes start to be depleted, leading  for  later times  ($t \sim 9$) to
a configuration where all the energy is concentrated only on the  $u^-_k$,
albeit they correspond to a small minority of the total number of
degrees-of-freedom.  Figure~\ref{fig3}(c) shows the flux of total energy as a
function of time. We observe a persistent constant positive flux corresponding
to inverse cascade of energy in the range $k\in [k_m,k_f]$.  This confirms that
also triads of Class II lead to a reverse energy cascade.  The energy is then
directly dissipated by the viscous effect which becomes substantial for the
highly energetic negative helical modes.  This is shown in fig.~\ref{fig3}(d),
where we compare the energy flux due to the nonlinear terms, 
\begin{align}
{\rm \Pi}_E(k)=\sum_{|\bk'|<k} {\hat {\bm u}}_{\bm k'}^* \cdot \hat{\bf N}_{\bk'},  
\label{eq:flux}
\end{align}
across a wavenumber $k$, where 
\begin{align}
\hat{\bf N}_\bk= \left(\mathbb{I}-\frac{\bk\bk}{k^2}\right)\left[\sum_{\bp+\bq=-\bk}(\bup\cdot\bq)\buq\right]
\end{align}
 is the nonlinear term in the Fourier space, and
the total molecular dissipation in the same Fourier interval: 
\begin{align}
{\rm D}(k)= 2\nu \sum_{|\bk'|<k} k'^2 E(k'). 
\end{align}
It should be noted that with this definition (\ref{eq:flux}) of energy flux, which has the
 opposite sign of what is commonly used, a positive/negative flux means
the presence of an inverse/direct energy cascade. Let us stress that the viscous 
contribution does not match exactly the
nonlinear transfer because the energy is still growing in time. Simulations for
the case where $k_f=4,6$ reach  a steady state earlier and  they show a much
better matching between the two contributions, see below panel (d) of
fig.~\ref{fig4}. Let us also notice that a sort of  $k^{-5/3}$ scaling is
observed in the inverse cascade regime as for the case when only Class I triads
are present~\cite{biferale2012}, at least up to the time when the condensate
does not become too energetic to spoil the scaling properties.

In fig.~\ref{fig4} we show the results from the case where  $k_m=6$. The main
interest to select this window  is that  in this way we can change the degree
of nonlocality of the triad geometry. In \cite{waleffe} it was argued that in
the  scaling regime triads of Class II should display either a forward or a
reverse energy transfer depending whether the ration between the smallest and
the medium wavenumber $v=k/p$ is larger or smaller than 0.278. If
we assume that the main energy transfer happens via a triad where two
wavenumbers fall in the forced range and the other belong to the negative
helical modes then we have $v=0.6$ for $k_m=6$ and $v=0.2$ for $k_m=2$.
As seen in fig.~\ref{fig4} we observe  an inverse energy
transfer also for $v=0.6$ contradicting the prediction made by~\cite{waleffe}. This is
probably due to the absence of any scaling regime for the  configuration of
forced and negative helical modes chosen here, as  shown by panel (a) and (b) of 
fig.~\ref{fig4}, and therefore our configuration does not satisfy the assumptions made in
\cite{waleffe}.  Figure~\ref{fig4}(d) shows the balance of ${\rm \Pi}_E(k)$ and ${\rm D}(k)$
for the wavenumbers $k\in[k_m,k_f]$ which confirms that negative helical modes
lose energy due to molecular  dissipation in such case. 

\subsection{Energy transfer for $k_m>k_f$}
In this second set of simulations we forced at $k_f\in[4,6]$ and kept the
negative helical modes only for larger wavenumbers,  $k_m=10$ and $k_m=16$. The
behavior of the growth of energy is similar to the cases of $k_m<k_f$ (see
fig.~\ref{fig5}). After the negative helical modes become energetic they
continue to accumulate energy and then reach a steady state by dissipating
energy directly via molecular viscosity.  However the dynamics of energy
transfer is entirely different from previous cases as seen in fig.~\ref{fig6}.
In fig.~\ref{fig6} (a) and (b) we show the spectrum for the total energy and
for the positive helical modes respectively.  As before, the difference between
the two gives the energy content in the negative helical modes.  In the
beginning we initialize the field at the forced scales and we observe a clear
inverse cascade of energy to large scales, shown by the energy spectra in
fig.~\ref{fig6} (a) and (b) and in the positive energy flux in
fig.~\ref{fig6}(c) at $t \sim 2.2$.  This transfer is due to the triads of
Class I.  Then, as soon as the negative helical modes become energetic enough,
the triads of Class III and Class IV take the lead and the energy flux is
reversed toward  the negative helical modes at scales smaller than the forced
ones from times $t \sim 4$ and larger.  It is interesting to observe that the
positive helical modes at large scales ($k<k_f$) also lose their energy by a
forward cascade, probably highly nonlocal. Figure~\ref{fig6}(c) shows the
evolution of the energy flux during the backward and forward regimes.
Panel (d) of the same figure compares the viscous contribution and the
nonlinear flux. The figure shows that in the late stationary regime the viscous
drag, induced by the high energy content of the negative helical modes, is
balanced with the nonlinear flux.  In this case we have a small-scales
condensate that adsorbs all energy flowing between modes at $k \sim k_f$ and $k
\sim k_m$. This is possibly due to the fact that positive helical modes
at $k>k_m$ do not receive energy from the negative helical modes at $k\sim k_m$
as they could only form triads of Class II which are responsible for inverse
energy transfer. 

\begin{figure*}[!htb]
\center
  \includegraphics[scale=0.35]{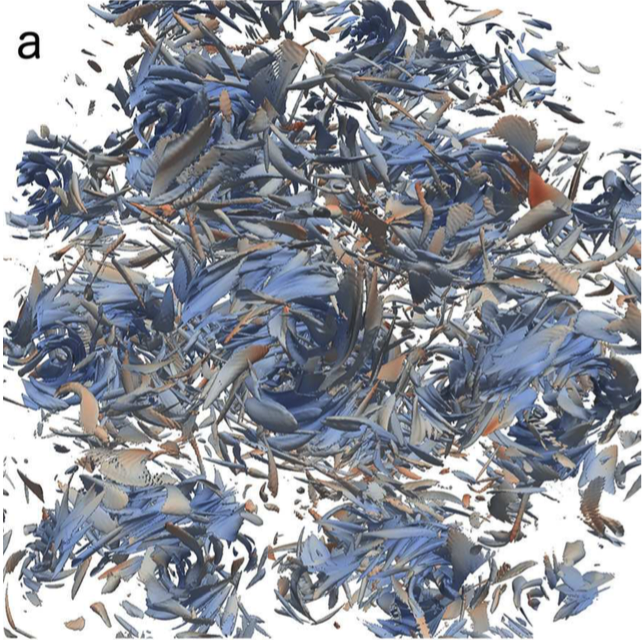}
  \includegraphics[scale=0.35]{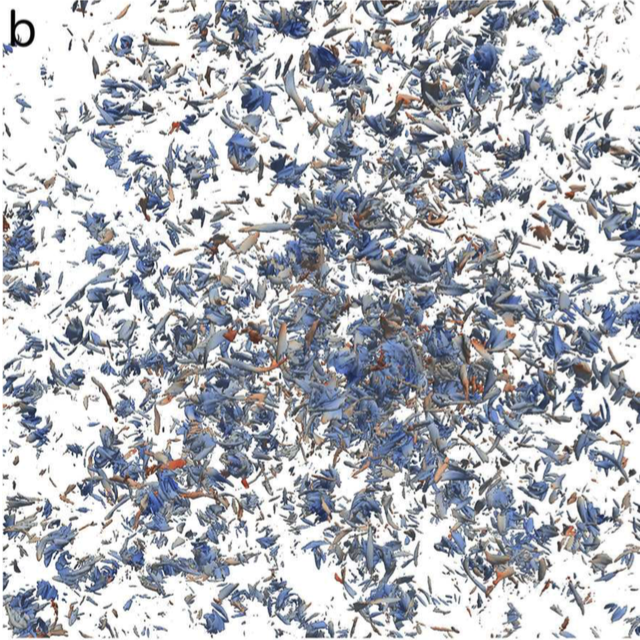}
  \caption{(Color online)  Iso-vorticity surfaces for: (a) $k_f=[10,12]$,
$k_m=4$, (b) $k_f=[4:6]$, $k_m=16$. Color palette is proportional to the
intensity of the helicity: red for high positive values ($\sim 10^3$) to blue
for high negative values ($\sim -10^3$).}
  \label{visual}
\end{figure*}

\subsection{Coherent structures}
As discussed in the previous sections, both experiments leads to a sort of
helical condensate concentrated on the wavenumbers where the negative helical
modes exists. This is a different way to produce (and stabilize) strong
nonlinear structures in Navier-Stokes equations with respects to the well known
case of two-dimensional turbulence~\cite{kraichnan67,boffetta,chertkov,cencini,clercx,laurie}.  A
visualization of the vorticity field where an inverse cascade of energy is
observed is shown in fig.~\ref{visual}(a). The presence of helical stable
structures is clearly detectable. In panel (b) of the same figure we show similar
small-scales condensates that populate the flow when $k_f\in[4,6]$ and
$k_m=16$. It would be interesting to understand if one
can highlight some universality properties of such configurations as done for
the two-dimensional case~\cite{laurie}. 
\section{Summary}

We have  performed several numerical simulations of a modified (decimated)
version of the three-dimensional Navier-Stokes equations by keeping only some
subsets of Fourier modes with different helical properties.  The aim is to
further understand the different roles played by triads with different helical
structures in the dynamics of the nonlinear energy transfer mechanism. We have
shown that as predicted in \cite{waleffe} there exist two classes (Class I and
Class II) of triads that transfer energy to large scales, i.e. which can
support an inverse cascade even in fully homogeneous and isotropic turbulence
(but not mirror symmetric). This result for Class I where all modes have the
same helical sign was already known~\cite{biferale2013,biferale2012}. The second class
(here called Class II) is made of triads where  helicity  is not globally
sign-definite.  The structure is such that the mode with the different helicity
is the one at the smallest wavenumbers. Hence, when the small-scales are
strongly helically-signed the forward energy transfer is depleted. The
existence of inverse cascade even when helicity is not positive-definite
contradicts the predictions based only on the absolute equilibrium in the
inviscid and unforced limit~\cite{herbert,kraichnan}.\\ 
By concentrating the negative helical modes at small scales (high wavenumbers) we showed
that as soon as triads of the other two classes (Class III and Class IV)
become competitive, they take the leadership in the energy transfer mechanisms
and the energy flux is reversed, reaching a more standard forward-cascade
regime. In both cases the energy is preferentially transferred to the minority
helical modes (here negative), leading to either a large-scale condensate or to a
small-scales condensate.  Our study further supports the idea that the direction
of the energy transfer in a turbulent flow might strongly be influenced by the
helicity distribution among  different scales~\cite{biferale2013,biferale2012,sahoo2015,stepanov,kessar}.

\section{acknowledgement}
We acknowledge useful discussions with F. Bonaccorso and funding from the
European Research Council under the European Union Seventh Framework
Programme, ERC Grant Agreement No 339032. Numerical simulations have been
partially supported by the INFN initiative INF14\_fldturb.


\begin{thebibliography}{99}
\bibitem{k41} A. N. Kolmogorov {\it Dokl. Akad. Nauk. SSSR} {\bf 32}, 19 (1941).
\bibitem{frisch} U. Frisch, {\it Turbulence: the legacy of A.N. Kolmogorov} (Cambridge University Press, Cambridge, UK,1995).
\bibitem{mininni} P. D. Mininni, A. Alexakis, and A. Pouquet, {\it Phys. Fluids} {\bf 21}, 015108 (2009).
\bibitem{deusebio} E. Deusebio and E. Lindborg, {\it J. Fluid Mech.} {\bf 755}, 654 (2014).
\bibitem{celani} A. Celani, S. Musacchio, and D. Vincenzi, {\it Phys. Rev. Lett.} {\bf 104}, 184506 (2010).
\bibitem{brandenburg} A. Brandenburg, {\it Astrophysical J.} {\bf 550}, 824 (2001).
\bibitem{fractal} U. Frisch, A. Pomyalov, I. Procaccia, and S. S. Ray {\it Phys. Rev. Lett.} {\bf 108}, 074501 (2012).
\bibitem{waleffe} F. Waleffe, {\it Phys Fluids A} {\bf 4}, 350 (1992).
\bibitem{constantin} P.  Constantin,  A.  Majda, {\it Commun.  Math.  Phys.}  {\bf 115}, 435 (1988).
\bibitem{biferale2013} L. Biferale, S. Musacchio, and F. Toschi, {\it J. Fluid Mech.} {\bf 730}, 309 (2013).
\bibitem{biferale-jstat} L. Biferale and E. S. Titi, {\it J. Stat. Phys.} {\bf 151}, 1089 (2013).
\bibitem{moffatt69} H. K. Moffatt,  {\it J. Fluid Mech.} {\bf 35}, 117 (1969).
\bibitem{moffatt92} H. K. Moffatt and A. Tsinober, {\it Annu. Rev. Fluid Mech.} {\bf 24}, 281 (1992).
\bibitem{brissaud} A. Brissaud, U. Frisch, J. Leorat, M. Lesieur, and M. Mazure, {\it Phys. Fluids} {\bf 16}, 1366 (1973).
\bibitem{laing} C.E. Laing, R. L. Ricca, and D. W. L. Summers, {\it Scientific Reports} {\bf 5}, 9224 (2015).
\bibitem{ditlevsen} P. D. Ditlevsen, {\it Phys. Fluids} {\bf 9}, 1482 (1997).
\bibitem{holm} D. D. Holm and R. M. Kerr, {\it Phys. Fluids} {\bf 19}, 025101 (2007).
\bibitem{biferaleh} R. Benzi, L. Biferale, R. M. Kerr, and E. Trovatore, {\it Phys. Rev. E} {\bf 53}, 3541 (1996).  
\bibitem{chen} Q. Chen, S. Chen, and G. L. Eyink, {\it Phys. Fluids} {\bf 15}, 361 (2003); 
\bibitem{chen2} Q. Chen, S. Chen, G. L. Eyink, and D. D. Holm, {\it Phys. Rev. Lett.} {\bf 90}, 214503 (2003).
\bibitem{dubrulle} E. Herbert, F. Daviaud, B. Dubrulle, S. Nazarenko, and A. Naso, {\it Europhys. Lett.} {\bf 100}, 44003 (2012).
\bibitem{biskamp} D. Biskamp, {\it Magnetohydrodynamic Turbulence} (Cambridge University Press, Cambridge, UK, 2003).
\bibitem{baer} J. Baerenzung, H. Politano, Y. Ponty, and A. Pouquet, {\it Phys. Rev. E} {\bf 77}, 046303 (2008).
\bibitem{mininni2010} P. D. Mininni and A. Pouquet, {\it Phys. Fluids} {\bf 22}, 035105 (2010).
\bibitem{biferale2012} L. Biferale, S. Musacchio, and F. Toschi, {\it Phys. Rev. Lett.} {\bf 108}, 164501 (2012). 
\bibitem{sahoo2015} G. Sahoo, F. Bonaccorsso, and L. Biferale, preprint, arXiv:1506.04906 (2015).
\bibitem{herbert} C. Herbert, {\it Phys. Rev. E} {\bf 89}, 013010 (2014).
\bibitem{kraichnan} R. H. Kraichnan, {\it J. Fluid Mech.} {\bf 47}, 525 (1971).
\bibitem{alexakis14} K Seshasayanan, S. J. Benavides, and A. Alexakis, {\it Phys. Rev. E} {\bf 90}, 051003 (2014).
\bibitem{alexakis15} K. Seshasayanan, A. Alexakis, preprint, arXiv:1509.02334 (2015).
\bibitem{moffatt14} H. K. Moffatt, {\it J. Fluid Mech.} {\bf 741}, R3 (2014).
\bibitem{kraichnan67} R. H. Kraichnan,  {\it Phys. Fluids} {\bf 10}, 1417 (1967).
\bibitem{boffetta} G. Boffetta and S. Musacchio, {\it Phys. Rev. E} {\bf 82}, 016307 (2010).
\bibitem{chertkov} M. Chertkov, C. Connaughton, I. Kolokolov, and V. Lebedev, {\it Phys. Rev. Lett.} {\bf 99}, 084501 (2007).
\bibitem{cencini} M. Cencini, P. Muratore-Ginanneschi, and A. Vulpiani, {\it Phys. Rev. Lett.} {\bf 107}, 174502 (2011).
\bibitem{clercx} H. J. H. Clercx, and G. J. F. van Heijst, {\it Appl. Mech. Rev.} {\bf 62}, 020802 (2009).
\bibitem{laurie} J. Laurie, G. Boffetta, G. Falkovich, I. Kolokolov, and V. Lebedev, {\it Phys. Rev. Lett.} {\bf 113}, 254503 (2014).
\bibitem{stepanov} R. Stepanov, E. Golbraikh, P. Frick, and A. Shestakov, preprint, arXiv:1508.07236 (2015).
\bibitem{kessar} M. Kessar, F. Plunian, R. Stepanov, and G. Balarac, preprint, arXiv:1509.02644 (2015).
\end{thebibliography}
\end{document}